# Rheology of Dilute Polymer Solutions with Time-Dependent Screening of Hydrodynamic Interactions

V. Lisy[1], J. Tothova[2], B. Brutovsky[2]

*The screening of hydrodynamic interactions (HI) essentially affects macroscopic properties of polymeric solutions. This screening depends not only on the polymer concentration, but has a dynamic nature. In the present work, a bead-spring theory is developed, in which this phenomenon is described for solutions of nonentangled polymer coils. The equation of motion for the beads of a test polymer is solved together with the Brinkman's equation for the solvent velocity that takes into account the presence of other coils in solution. The time correlation functions for the polymer normal modes are found. A tendency to the screening of HI is demonstrated on the coil diffusion as well as on the relaxation of its internal modes. With the growing concentration of the coils they both show a transition to the exact Rouse behavior. The viscosity of the solution and some other observable quantities are calculated. As the time increases, the time-dependent quantities change their behavior from the Rouse regime through the Zimm one again to the Rouse dynamics at long times.*

## I. Introduction

Rheology of polymer solutions represents a cross-disciplinary field, using wide spectra of theoretical tools from physics and chemistry. For physicists, understanding the configuration and dynamics of long polymer chains has been a significant source of problems within statistical physics from the 1950's onwards. One of the reasons why physicists were drawn to the problem is the universality of polymer properties [1]. Within the time and length scales much exceeding the atomic ones, universal theories have been built, well describing the main features in the polymer behavior, insensitive to the details of the chemistry of the chains. Among these theories the most popular are the Rouse and Zimm models, in which the polymer is represented as a chain of beads under Brownian motion [2]. The present work was inspired by the difficulties that still exist between these models and experiments. For example, one can find different results for the viscosity of dilute polymer solutions (see [1-4] and the citations there), the observed monomer motion in single polymer chains cannot be explained by the available theories [5, 6], there are problems with the description of the dynamic light scattering from polymer solutions [7, 8], and others [9]. Although the Rouse and Zimm models give a good qualitative base for the polymer solution rheology, some of the existing quantitative discrepancies with the experiments are not resolved for decades and new problems appear. An example is the computer simulation study [10], where the screening of hydrodynamic interactions (HI) in semidilute polymer solutions was investigated. It is well known that, due to this screening, with the increase of their concentration the polymers in solution change their behavior from the Zimm- to the Rouse-like one [2]. It has been found in [10] that this process is not only concentration dependent but has a dynamic character, i.e., it changes with the time. Based on our previous generalization of the Rouse-Zimm theory [11, 12], this observation can be explained theoretically in a natural way. Our approach to the problem can be summarized as follows. First, as distinct from the traditional use of the Zimm model for polymers in solution (this corresponds to strong HI and is in the theory characterized by the infinite draining parameter $h$ [2]), we consider this model just as a special case of a more general theory with a finite $h$. Due to this, both the diffusion coefficient of the whole polymer coil and the relaxation rates of its internal modes are sums of Rouse and Zimm contributions, the former one being usually omitted in the interpretation of experimental data. Secondly, the internal modes have their own draining parameter $h(p)$, which depends on the mode number $p$ = 1, 2, ... When $p$ increases, $h(p)$ decreases and beginning from some $p$ all the higher internal modes become the Rouse modes even if the whole polymer is predominantly of the Zimm type. Third, the type of the dynamics depends also on the time. Consider, for example, the mean square displacement (MSD) of a polymer segment. At sufficiently short times its dynamics in a flexible polymer is Rouse-like and with the growing time it changes to the Zimm-type dynamics (the relative contribution of the modes with small draining parameters decreases, or, in other words, the influence of the HI increases) [6]. This crossover cannot be described coming from approximate expressions, according to which the MSD is given by the simple $t^{1/2}$ or $t^{2/3}$ laws for the "Rouse" or "Zimm" polymers, respectively [13, 14]. These laws have been obtained assuming the continuous distribution of internal modes with respect to $p$. However, this assumption is true only in a restricted time domain and can lead to incorrect interpretation of experimental data [5]. Using the discussed approach, we have recently [15, 6] interpreted the fluorescence correlation spectroscopy measurements on individual DNA molecules and showed that the original interpretation [5] of the data was misleading. When the motion of a macromolecule is affected by other coils in the solution, one more problem arises: it is necessary to take into account the forces with which the coils act on the solvent and thus hamper its flow. For nonentangled polymers it can be done using the Debye-Bueche-Brinkman theory for the flow between obstacles [16, 17]. The friction factor $f$ on the polymer chain during its motion can be determined from the Einstein relation $D = k_BT/f$, where $D$ is the coil diffusion coefficient. By this way, in the next section the generalized Rouse-Zimm equation for the position vectors of the beads and the Oseen tensor describing the velocity field of the solvent due to perturbations are built. The described approach then allows us to obtain new results on the diffusion of the polymer coils and their internal motion and to find the related observable quantities, such as the viscosity functions of the polymer solution or the relaxation modules. With the growing concentration the Zimm contribution to these quantities disappears. Our theory thus describes in a simple manner the tendency to the transition between the Zimm and *exact* Rouse behavior of the polymer. Along with the concentration dependence, the dynamic nature of the screening of HI is revealed: as the time increases, the time-dependent quantities, such as the bead MSD, change their behavior from the Rouse regime through the Zimm one again to the Rouse dynamics.

## II. HYDRODYNAMICS OF POLYMER SOLUTIONS

We choose one of the polymers in solution as a "test" polymer [2]. The equation of motion of its $n$th bead is

$$M\frac{d^2\vec{x}_n(t)}{dt^2} = \vec{f}_n^{\,fr} + \vec{f}_n^{\,ch} + \vec{f}_n \,. \qquad (1)$$

Here, $\vec{x}$ is the position vector of the $n$th bead from $N$ ones mapping the polymer, $M$ is the bead mass, $\vec{f}_n^{\,ch}$ the force with which the neighboring beads act on the $n$th bead, $\vec{f}_n$ is the random force due to the motion of the molecules of solvent, and $\vec{f}_n^{\,fr}$ is the Stokes friction force on the bead during its motion in the solvent [18],

$$\vec{f}_n^{\,fr} = -\xi\left[\frac{d\vec{x}_n}{dt} - \vec{v}(\vec{x}_n)\right], \qquad (2)$$

where $\vec{v}(\vec{x}_n)$ denotes the velocity of the solvent in the place of the $n$th bead due to the motion of other beads. The friction coefficient is $\xi = 6\pi\eta b$ ($b$ is the bead radius and $\eta$ the solvent viscosity). This expression holds in the case of steady flow. In a more general case taking into account the hydrodynamic memory [11, 19, 20] the force (2) must be replaced by the Boussinesq force and (1) has to be solved together with the nonstationary hydrodynamic equations for the macroscopic velocity of the solvent. To take into account the presence of other polymers in solution, we use the Brinkman-Debye-Bueche [15, 16]) theory in which the polymer is considered as a porous medium. In our approach all the solution is such a medium permeable to the solvent flow. Then in the right hand side of the Navier-Stokes equation a term $-\kappa^2\eta\vec{v}$ has to be added, where $1/\kappa^2$ is the solvent permeability. This term corresponds to the average value of the force acting on the liquid in an element of volume d$V$, provided the average number of polymers in solution per d$V$ is $c$; then $\kappa^2\eta = cf$, where $f$ is the friction factor on one polymer chain. Thus,

$$\rho\frac{\partial\vec{v}}{\partial t} = -\nabla p + \eta\Delta\vec{v} - \kappa^2\eta\vec{v} + \vec{\varphi}\,. \qquad (3)$$

Here, $\rho$ is the solvent density, $p$ is the pressure, and $\vec{\varphi}$ is the density of the force from the polymer beads on the liquid [18],

$$\vec{\varphi}(\vec{x}) = -\sum_n \vec{f}_n^{\,fr}(\vec{x}_n)\delta(\vec{x}-\vec{x}_n)\,. \qquad (4)$$

Solving this equation is a difficult problem since the polymer coils are moving. However, in the first approximation, small and slow changes of the concentration $c(t)$ around its equilibrium value can be neglected. The beads are much more mobile than the whole coils of long polymer chains ($N >> 1$). This is seen comparing the diffusion coefficient of one bead, $D_b = k_BT/(6\pi b\eta)$, with that of the coil in the Zimm ($D_Z = 8k_BT/[3(6\pi^3N)^{1/2}a\eta]$) or Rouse ($D_R = k_BT/(6\pi Nb\eta)$) limits [2] ($a$ is the mean square distance between the beads along the chain). In the latter case $D_R/D_b = 1/N$, and for the Zimm polymers $D_Z/D_b \approx 3.7b/(a\sqrt{N})$. The motion of the solvent created by the motion of beads is thus much faster than the motion of the coils, which determines the changes of $c(t)$.

Eqs. (1) - (4) describe the motion of one bead in the solvent, when the obstacles (other coils) influence the solvent flow. This problem can be transformed to that first solved in [21] (for later works see also [20, 11]). The velocity field can be in the Fourier representation in the time written as follows:

$$v_\alpha^\omega(\vec{r}) = \int d\vec{r}'\sum_\beta H_{\alpha\beta}^\omega(\vec{r}-\vec{r}')\varphi_\beta^\omega(\vec{r}')\,. \qquad (5)$$

Here, the analog of the Oseen tensor is

$$H_{\alpha\beta}^\omega(\vec{r}) = A\delta_{\alpha\beta} + B\frac{r_\alpha r_\beta}{r^2}\,, \qquad (6)$$

where

$$A = \frac{1}{8\pi\eta r}\left[e^{-y} - y\left(\frac{1-e^{-y}}{y}\right)''\right],$$

$$B = \frac{1}{8\pi\eta r}\left[e^{-y} + 3y\left(\frac{1-e^{-y}}{y}\right)''\right], \qquad (7)$$

$y = r\chi$, $\chi^2 = \kappa^2 - i\omega\rho/\eta$, and the prime means the differentiation with respect to $y$. In the particular case $\omega = 0$ and for permeable solvents when $\kappa = 0$, Eqs. (6) and (7) coincide with the result of Zimm [2]. Using this solution, a generalization of the Rouse-Zimm equation has been obtained from the equation of motion [11]. The preaveraging of the Oseen tensor over the equilibrium (Gaussian) distribution of the beads [2, 18] gives

$$\left\langle H^{\omega}_{\alpha\beta nm}\right\rangle_0 = \delta_{\alpha\beta} h^{\omega}(n-m), \quad \vec{r}_{nm} \equiv \vec{x}_n - \vec{x}_m, \qquad (8)$$

with

$$h^{\omega}(n-m) = \left(6\pi^3 |n-m|\right)^{-1/2} (\eta a)^{-1}$$

$$\times\left[1 - \sqrt{\pi} z \exp\left(z^2\right)\mathrm{erfc}(z)\right], \quad z \equiv \chi a\sqrt{|n-m|/6}.$$

Then, in the continuum approximation with respect to the variable $n$ [2, 18], the new Rouse-Zimm equation contains only the diagonal terms [11],

$$-i\omega\vec{x}^{\omega}(n) = \frac{1}{\xi}\left[\frac{3k_BT}{a^2}\frac{\partial^2 \vec{x}^{\omega}(n)}{\partial n^2} + M\omega^2\vec{x}^{\omega}(n) + \vec{f}^{\omega}(n)\right]$$

$$+\int_0^N dm\, h^{\omega}(n-m)\left[\frac{3k_BT}{a^2}\frac{\partial^2 \vec{x}^{\omega}(m)}{\partial m^2}\right.$$

$$\left. + \; M\omega^2\vec{x}^{\omega}(m) + \vec{f}^{\omega}(m)\right]. \qquad (9)$$

It is solved with the help of the Fourier transformation (FT) in $n$, $\vec{x}^{\omega}(n) = \vec{y}^{\omega}_0 + 2\sum_{p=1}^{\infty}\vec{y}^{\omega}_p \cos(\pi n p/N)$, taking into account the conditions at the ends of the chain, $\partial\vec{x}(t,n)/\partial n = 0$ at $n = 0, N$. The inverse FT yields

$$\vec{y}^{\omega}_p = \vec{f}^{\omega}_p\left[-i\omega\Xi^{\omega}_p - M\omega^2 + K_p\right]^{-1}, \qquad (10)$$

where

$$\Xi^{\omega}_p = \xi\left[1 + \left(2-\delta_{p0}\right)N\xi h^{\omega}_{pp}\right]^{-1}, K_p = 3k_BT\left(\frac{\pi p}{Na}\right)^2 \qquad (11)$$

$p = 0, 1, 2, \ldots$, and the Oseen matrix $h^{\omega}_{pp}$ reads [11],

$$h^{\omega}_{00} = \frac{2}{\sqrt{6N}\pi\eta a\chi_{\omega}}\left[1 - \frac{2}{\sqrt{\pi}\chi_{\omega}} + \frac{1}{\chi^2_{\omega}}\left(1 - \exp\chi^2_{\omega}\,\mathrm{erfc}\chi_{\omega}\right)\right],$$

$$h^{\omega}_{pp} = \frac{1}{\pi\eta a\sqrt{3\pi N p}}\frac{1+\chi_p}{1+\left(1+\chi_p\right)^2}, \qquad p = 1, 2, \ldots \qquad (12)$$

where $\chi_{\omega} = \sqrt{N/6}\chi a$ and $\chi_p = \sqrt{N/(3\pi p)}\chi a$. Using the fluctuation-dissipation theorem [24] or the properties of the random forces $\vec{f}^{\omega}_p$ [25], the time correlation functions for the normal modes are determined as

$$\psi_p(t) = \left\langle y_{\alpha p}(0)\, y_{\alpha p}(t)\right\rangle \qquad (13)$$

$$= \frac{k_BT}{\left(2-\delta_{p0}\right)\pi N}\int_{-\infty}^{\infty} d\omega \cos\omega t \frac{\mathrm{Re}\,\Xi^{\omega}_p}{\left|-i\omega\Xi^{\omega}_p - M\omega^2 + K_p\right|^2}.$$

In the stationary limit $\omega = 0$ so that $\chi = \kappa$. Then the preaveraged Oseen tensor (6) is

$$\left\langle H^{\omega}_{\alpha\beta}\right\rangle_0 = \frac{\delta_{\alpha\beta}}{6\pi\eta}\left\langle\frac{e^{-\chi r}}{r}\right\rangle_0, \qquad (14)$$

and the quantity $1/\kappa$ can be for small $\kappa r$ considered as a screening length. Let us first focus on the motion of the center of inertia of the polymer.

### *II.1. Diffusion of the Polymer Coil*

For an individual polymer and $p = 0$ in Eq. (13),

$$\psi_0(0) - \psi_0(t) = Dt, \qquad (15)$$

with the diffusion coefficient $D = D_R + D_Z$ ($R$ and $Z$ stay for the Rouse and Zimm limits given above). Now instead of Eq. (12) we have $h^0_{00}$ with $\chi_0 = \kappa R_G$ ($R_G = \sqrt{Na^2/6}$ is the gyration radius), $D$ depends on the concentration of the coils $c$,

$$D(c) = D_Z(c) + D_R, \quad D_Z(0) = D_Z, \qquad (16)$$

and consists of the Rouse (independent on the presence of other polymers) and the Zimm contributions. The latter one can be expressed in the form

$$D_Z(c) = D_Z(0) f(c), \qquad (17)$$

where $f(c)$ is a universal function for every polymer:

$$f(c) = \frac{3\sqrt{\pi}}{4\chi_0}\left[1 - \frac{2}{\sqrt{\pi}\chi_0} - \frac{1}{\chi^2_0}\left(\exp\chi^2_0\,\mathrm{erfc}\chi_0 - 1\right)\right]. \qquad (18)$$

The dependence of the permeability on $c$ can be estimated as follows. The friction coefficient in the quantity $\kappa^2 = cf/\eta$ from Eq. (3) can be determined from the Einstein relation $D = k_BT/f$. Then

$$\kappa^2 = \frac{27\sqrt{\pi}}{16}\frac{\tilde{c}}{R^2_G}\left(1 + \frac{3}{4\sqrt{2}h}\right)^{-1}. \qquad (19)$$

The values of $\kappa$ and $\chi_0$ depend on the draining parameter $h = 2\sqrt{3N/\pi}b/a$ (if $h >> 1$, the coil dynamics is of the Zimm type, for $h << 1$ we have the free-draining Rouse limit). The quantity $\tilde{c} \equiv 4\pi R_G^3 c/3$ denotes the number of polymers per the volume of a sphere with the radius $R_G$. With the increase of $c$ the Zimm term decreases and for large $c$ (small permeability $\kappa$; when $\chi_0 >> 1$) it becomes ~ $1/\sqrt{c}$ ,

$$D_Z(c) \approx \frac{2k_BT}{\pi\eta Na^2}\frac{1}{\kappa}. \quad (20)$$

The realistic case of small $c$ corresponds to $\chi_0 = \kappa R_G << 1$ when

$$D_Z(c) = k_BTh_{00}^0(c) = D_Z(0)\left(1 - \frac{3}{8\sqrt{\pi}}\kappa R_G + ...\right). \quad (21)$$

The $c$-dependent correction to $D_Z(0)$ is proportional to $\sqrt{c}$ and differs from other results (e.g. [26], where this correction is ~ $c$). When the polymer is free, the type of its diffusion depends only on the draining parameter $h$. With the growing $c$, the polymer changes its behavior to the diffusion with the exactly Rouse coefficient $D_R$.

### II.2. Dynamics of Internal Modes

In the stationary case and at $\kappa = 0$ the diagonal elements of the Oseen matrix are [2]

$$h_{pp}^0(0) = \left(12\pi^3 Np\right)^{-1/2}(\eta a)^{-1}. \quad (22)$$

Now $h_{pp}^0(c)$ from (13) depends on $c$. The internal modes relax exponentially,

$$\psi_p(t) = \frac{k_BT}{2NK_p}\exp\left(-|t|/\tau_p\right) \quad (23)$$

and their relaxation rates consist of the Rouse contribution and the $c$-dependent Zimm part,

$$\frac{1}{\tau_p(c)} = \frac{1}{\tau_{pR}} + \frac{1}{\tau_{pZ}(c)}, \quad (24)$$

where $\tau_{pR}$ and $\tau_{pZ}(0) \equiv \tau_{pZ}$ are given by [2]

$$\tau_{pR} = \frac{2N^2a^2b\eta}{\pi k_BTp^2}, \quad \tau_{pZ} = \frac{\left(\sqrt{N}a\right)^3\eta}{\left(3\pi p^3\right)^{1/2}k_BT}, \quad (25)$$

and

$$\tau_{pZ}(c) = \frac{1}{2}\tau_{pZ}(0)\frac{1+\left(1+\chi_p\right)^2}{1+\chi_p}, \quad (26)$$

which at $c \to 0$ behaves as

$$\tau_{pZ}(c) = \tau_{pZ}(0)\left(1 + \frac{N}{6\pi p}\kappa^2a^2 - ...\right), \quad (27)$$

and as $c \to \infty$ one has

$$\tau_{pZ}(c) \approx \frac{1}{2}\tau_{pZ}(0)\chi_p = \frac{\left(Na^2\right)^2\eta}{6\pi k_BTp^2}\kappa. \quad (28)$$

### II.3. Steady State Viscosity

Viscosity is the most important property that determines the flow characteristics of the fluid. Using the above calculated relaxation times $\tau_p$ of the polymer internal modes, the steady state viscosity of the solution can be calculated from the formula [9]

$$\eta(c) = \eta + \frac{1}{2}k_BTc\sum_{p=1}^{\infty}\tau_p(c). \quad (29)$$

In the Rouse limit we obtain the known result [2] $\eta(c) = \eta + \pi N^2a^2bc\eta/6$. In the Zimm case at low concentrations

$$\frac{\eta(c)-\eta}{\eta} = \frac{c}{2\sqrt{3\pi}}\left(Na^2\right)^{3/2}\sum_{p=1}^{\infty}p^{-3/2}\left[1 + \frac{\sqrt{6\pi}}{16p}c\left(Na^2\right)^{3/2} + ...\right]$$

$$= 0.425c\left(Na^2\right)^{3/2}\left[1 + 0.140c\left(Na^2\right)^{3/2} + ...\right], \quad (30)$$

where the first term corresponds to the known formula [2]). In our theory, the most general expression for the viscosity is

$$\frac{\eta(c)-\eta}{\eta c} = \frac{N^2a^2b}{\pi}\sum_{p=1}^{\infty}\frac{1}{p^2}\left(1 + \frac{2h}{\sqrt{p}}\frac{1+\chi_p}{1+\left(1+\chi_p\right)^2}\right)^{-1} \quad (31)$$

At very low concentrations when $\chi_p << 1$, we have for the so called intrinsic viscosity

$$[\eta]_h = \frac{\eta(c)-\eta}{\eta c} = \frac{1}{\pi}N^2a^2b\sum_{p=1}^{\infty}\frac{1}{p^2}\left(1 + \frac{h}{\sqrt{p}}\right)^{-1}. \quad (32)$$

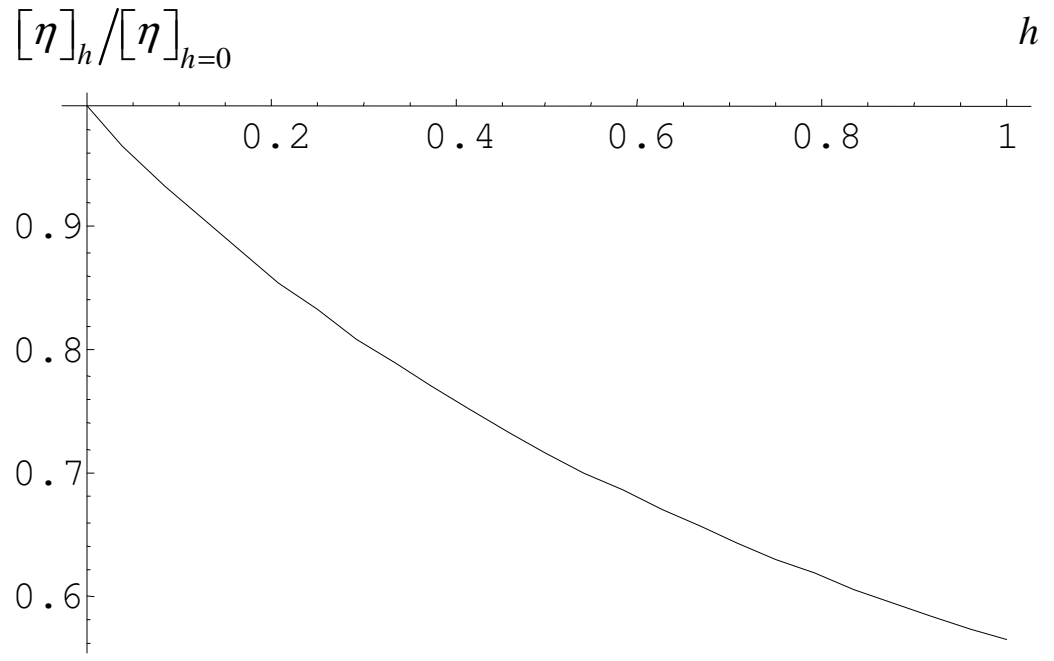

**Fig. 1.** Viscosity normalized to its Rouse limit as a function of $h < 1$, when the polymer is assumed to be the Rouse one.

Due to the dependence on $h$ the difference between $[\eta]_h$ and the classical results can be notable. For a Rouse polymer with small $h$ this is illustrated by Fig. 1. In Fig. 2, $[\eta]_{h >> 1}$ ~ $1/h$ is the intrinsic viscosity of the Zimm polymer, for which $[\eta(c) - \eta]/\eta = 2.61N^2a^2bc/(\pi h)$. It is seen that even for $h \approx 10$ the difference from the Zimm viscosity is ~ 20%.

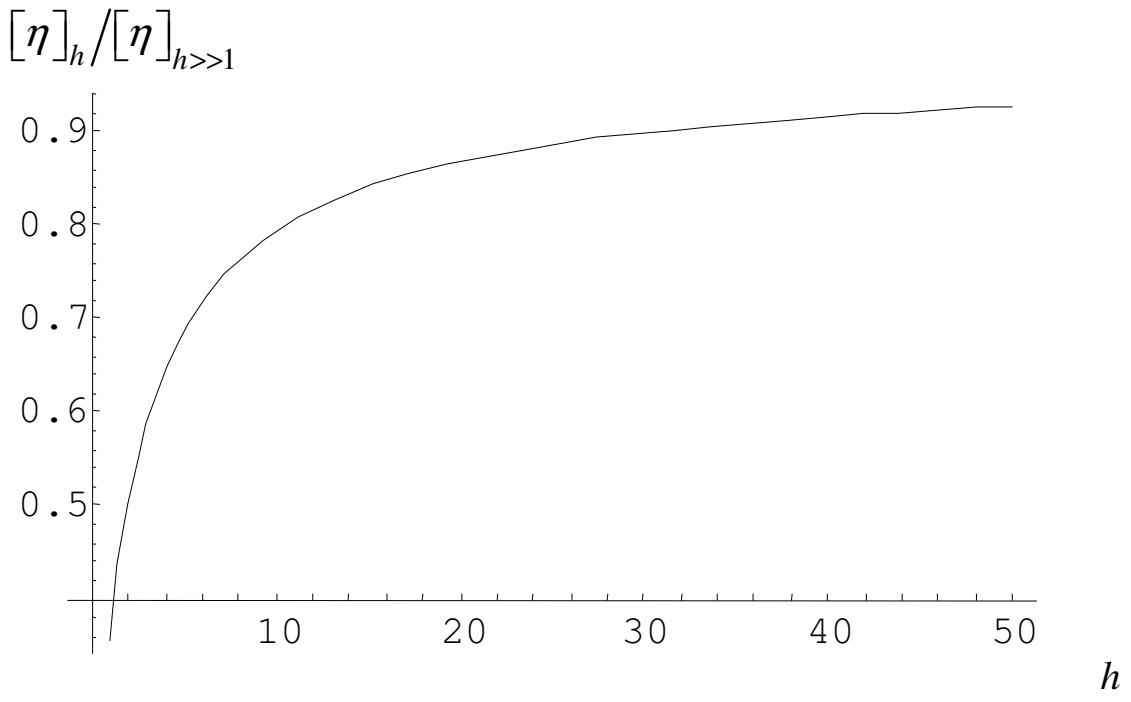

**Fig. 2.** The same as in Fig. 1 for large $h$ (the Zimm polymer).

Using the above results, the Huggins coefficient $k_H$ [2], which is one of the most often determined quantities in viscosimetry measurements, can be found. From the general expression for the viscosity (31), the Huggins equation is

$$\frac{\eta(c)-\eta}{\eta c} = [\eta]\left(1+k_H[\eta]c+...\right), \qquad (33)$$

where

$$k_H = \pi h\left(1+\frac{4\sqrt{2}h}{3}\right)^{-1}\left[\sum_{p=1}^{\infty}\frac{1}{p^2}\left(1+\frac{h}{\sqrt{p}}\right)^{-1}\right]^{-2}$$

$$\times\sum_{p=1}^{\infty}\frac{1}{p^{7/2}}\left(1+\frac{h}{\sqrt{p}}\right)^{-2}. \qquad (34)$$

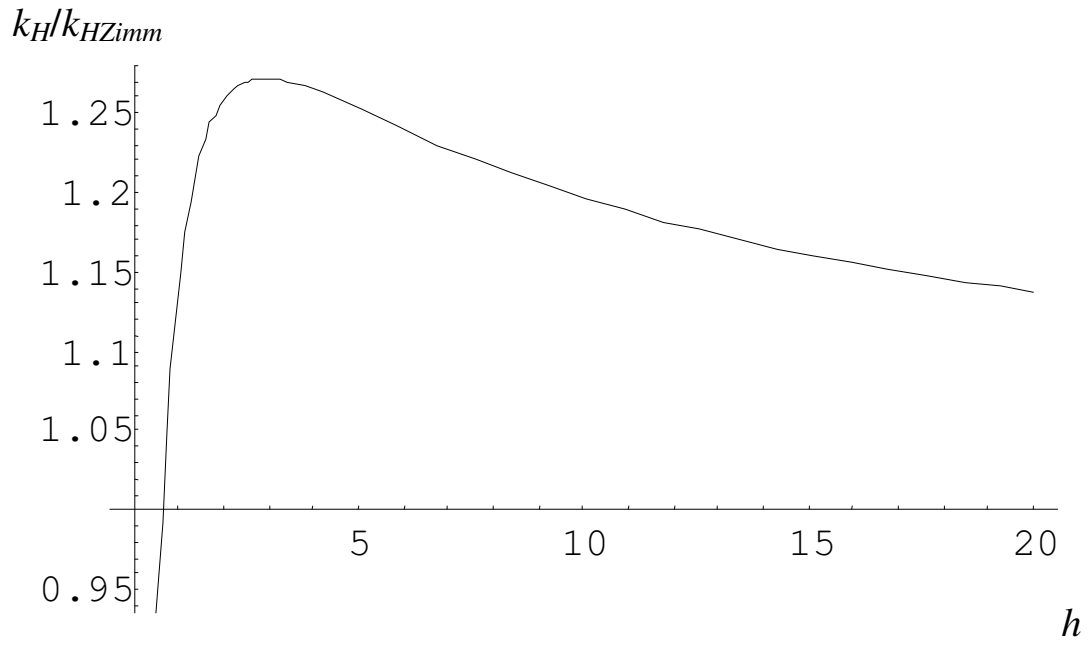

**Fig. 3.** Huggins coefficient normalized to its Zimm limit.

In Fig. 3, the Huggins coefficient related to its Zimm limit is shown. It is seen that with the growing $h$, $k_H$ slowly approaches $k_{HZimm}$. The difference is significant in a broad region of $h$.

For large $h$ (the Zimm case) we find

$$[\eta]_\infty = \frac{N^{3/2}a^3}{2\sqrt{3\pi}}\sum_{p=1}^{\infty}p^{-3/2} = 3\sqrt{\frac{2}{\pi}}R_G^3\zeta\left(\frac{3}{2}\right) \approx 6.253\, R_G^3, \qquad (35)$$

where $\zeta$ is the Riemann zeta function. In this case

$$k_{HZimm} = 3\pi 2^{-5/2}\zeta(5/2)\zeta^{-2}(3/2) \approx 0.3275. \qquad (36)$$

Note that in our work [12] the factor 1/2 is missing in the expression for $k_{HZimm}$. This result differs from the known results (e.g., Doi and Edwards [2] give the value 0.757, $k_{HZimm}$ = 0.6949 in [4], etc.; see the discussion in [3, 4]). The works [27, 28] possess the viscosity, which is inconsistent with the Kirkwood and Riseman [29] theory and gives the hydrodynamic screening even for infinitely dilute solutions. The work [28] suggests that the screening cannot be described if the preaveraging approximation is employed for the HI; as shown here, this is not true. Finally, in the opposite Rouse limit when $h \to 0$, $k_H$ approaches zero as $k_H \approx \pi h\zeta(3.5)\zeta^2(2) \approx 1.3\, h$.

### II.4. The Relaxation Modulus

The relaxation modulus $G$ determines the shear stress at shear flows with the velocity $\upsilon_x(\vec{r},t) = \zeta(t) r_y$, $\upsilon_y = \upsilon_z = 0$ [2, 18],

$$\sigma_{xy}(t) = \eta\zeta(t) + \int_{-\infty}^{t} dt' G^p(t-t')\zeta(t')$$

$$= \int_{-\infty}^{t} dt' G(t-t')\zeta(t'). \qquad (37)$$

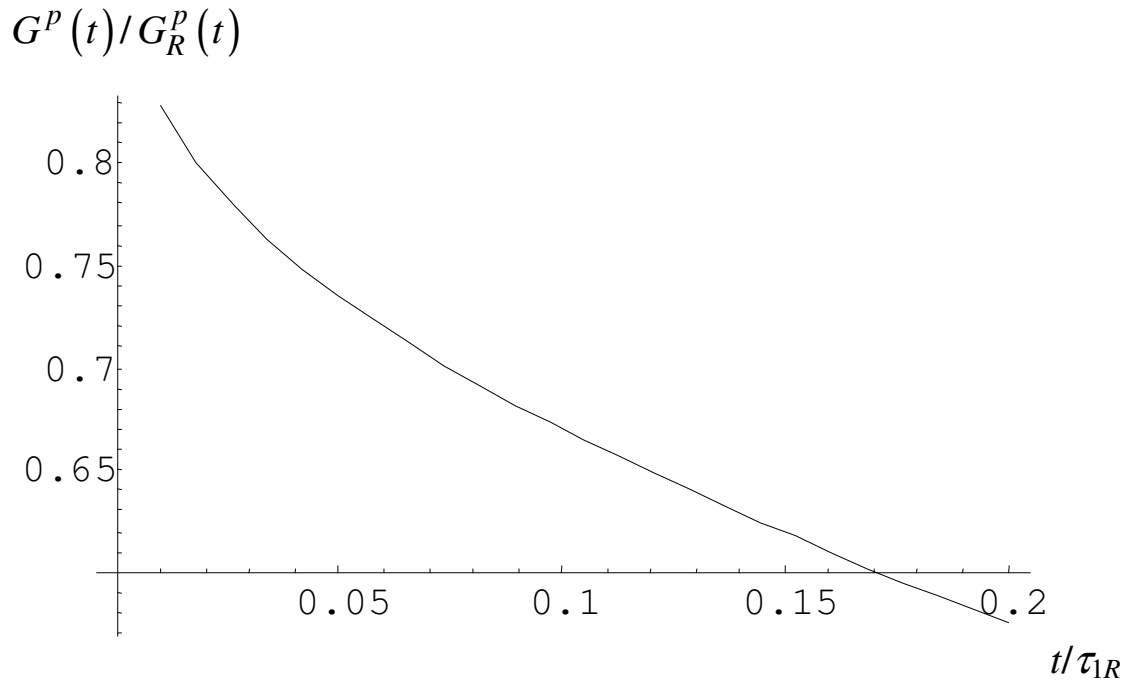

**Fig. 4.** Relaxation modulus $G^p$ as a function of $t$ at $h$ = 1.

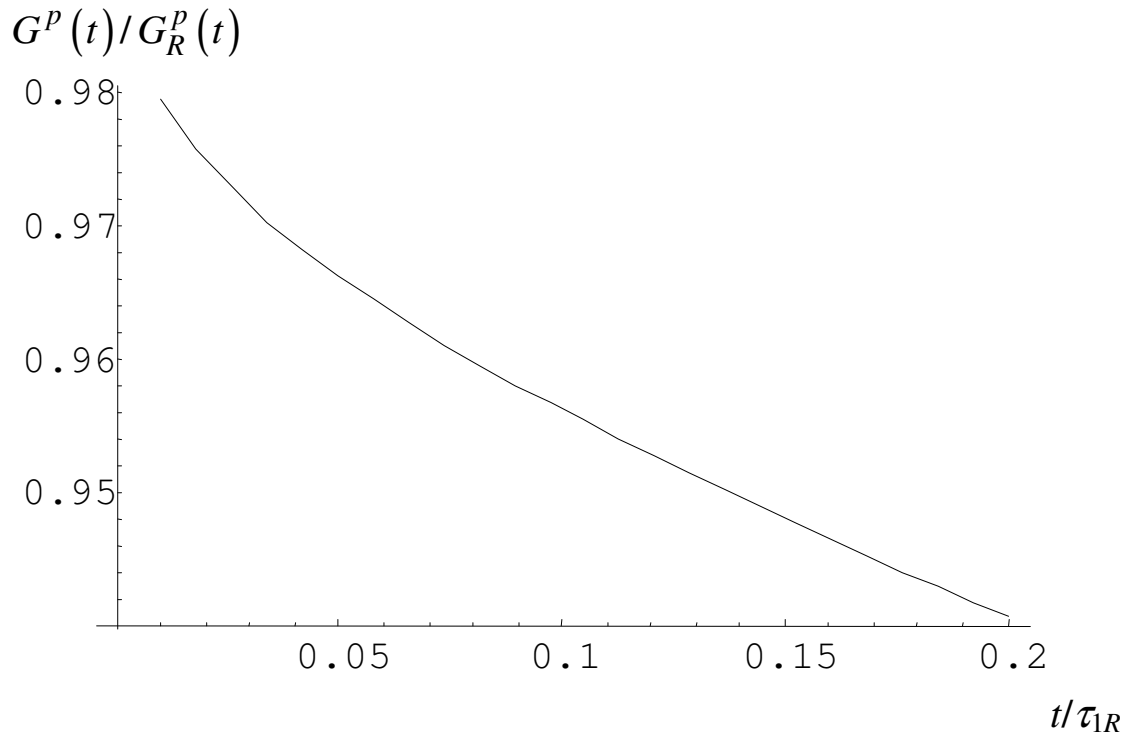

**Fig. 5.** The same as in Fig. 4. at $h$ = 0.1.

Having solved the chain dynamics, the modulus $G$ is calculated from the equation

$$G^p(t) = k_B T c \sum_{q=1}^{\infty} \exp\left(-2t/\tau_q\right). \qquad (38)$$

Figs. 4 and 5 show $G$ at $c$ = 0, related to the Rouse model. With the growing $t$, the difference from the Rouse result becomes significant even at small $h$. So, when $t/\tau_{1R} \approx 10$, even for $h$ = 1/100 the difference between $G^p$ and its Rouse limit is about 20%.

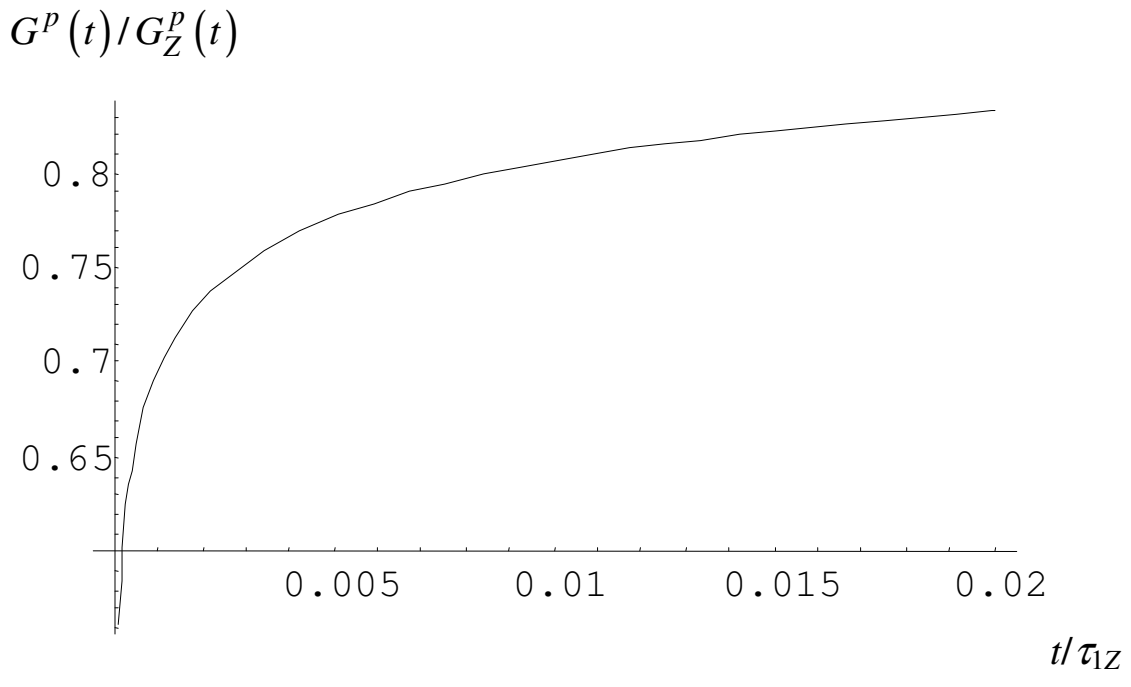

**Fig. 6.** Short-time behavior of the relaxation modulus related to its Zimm limit. The draining parameter is $h$ = 10.

Fig. 6 illustrates the difference of $G^p(t)$ from the pure Zimm behavior. At very short times this difference is significant even for large $h$. With the increase of $t$, $G^p$ becomes closer to its Zimm limit; however, a transition to the Rouse behavior at long times is observed, as shown in Fig. 7. For the chosen $h$ = 10, the difference from the Zimm modulus is always larger than 10%.

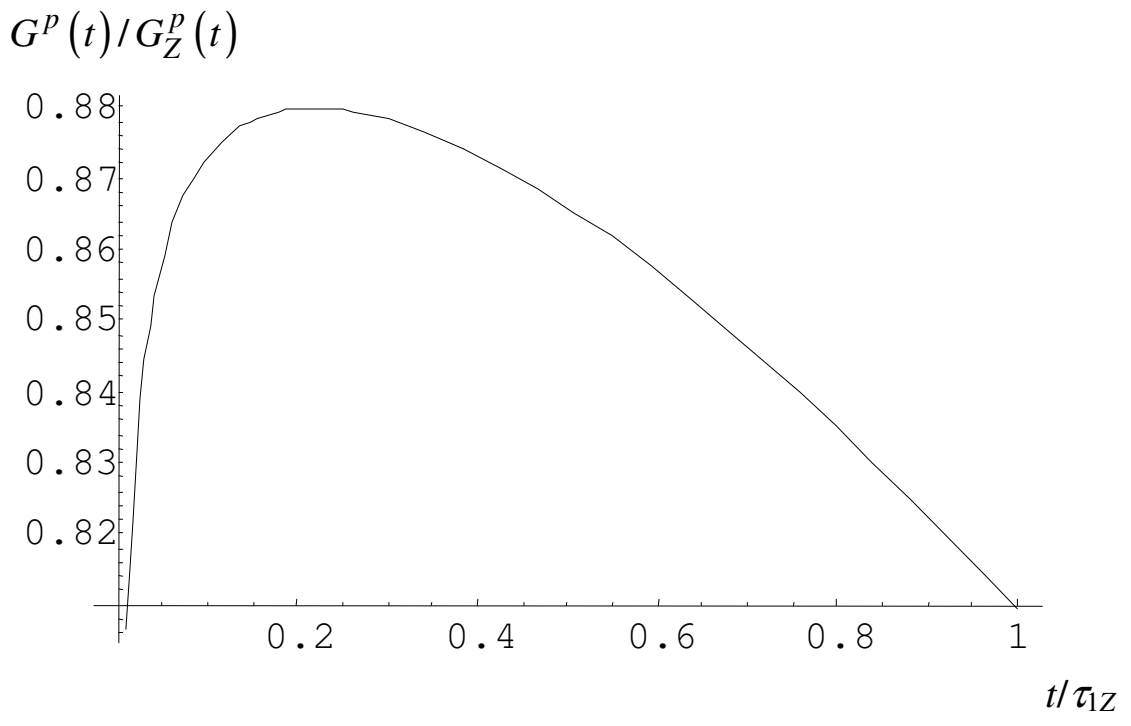

**Fig. 7.** The same as in Fig. 6 at longer times.

### II.5. Monomer MSD

Similar results can be obtained for other quantities, like the complex modulus, the dynamic structure factor of the test polymer [8], or the MSD of a monomer within an isolated coil [5, 6].

The internal modes of the polymer contribute to the MSD of its end monomer as follows [6]:

$$\left\langle r^2(t)\right\rangle_{\text{int}} = \frac{4Na^2}{\pi^2}\sum_{p=1}^{\infty}\frac{1}{p^2}\left[1-\exp\left(-\frac{t}{\tau_p(c)}\right)\right]. \qquad (39)$$

The numerical calculations using this expression are given in Fig. 8. We relate the Rouse MSD (at $h$ = 0) to the Rouse-Zimm MSD at $h$ = 10, to show how this function changes depending on the time at a relatively low concentration (one coil in the volume $10\times 4\pi R_G^3/3$). It is seen that at long times the behavior of the polymer, which was initially predominantly of the Zimm type, changes to the Rouse-like type.

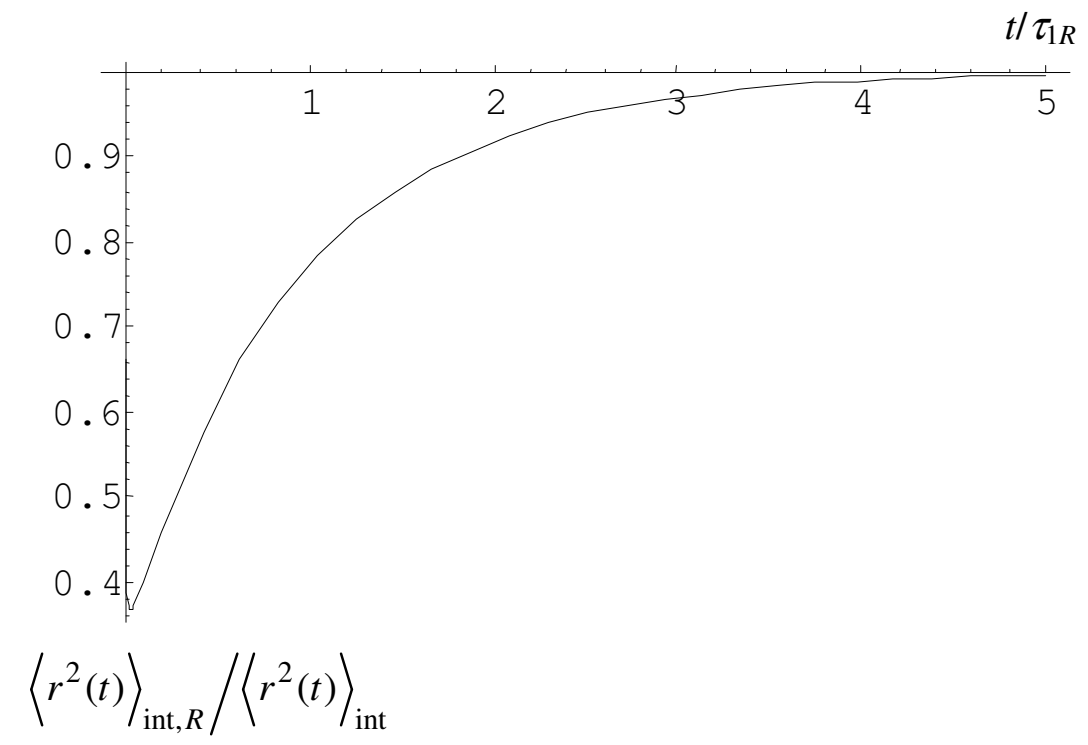

**Fig. 8.** Relation of the Rouse MSD (at $h$ = 0) to the Rouse-Zimm MSD at $h$ = 10 and $\tilde{c} = 0.1$ as a function of $t/\tau_{1R}$.

Some more calculations are given in [6] for a single polymer coil and in [8], where the influence of other coils is considered. These results can be summarized as follows: every polymer at very short times at any $c$ behaves as the Rouse one since the HI does not yet affect the dynamics. At longer times, the HI takes effect and the polymer begins to move in the Zimm regime. Then, due to the screening of HIs, the polymer behavior turns again to the Rouse-like one.

## III. Conclusion

The properties of complex polymeric systems cannot be comprehended without understanding the dynamics of a single polymer in well defined conditions, such as in dilute solutions of nonentangled polymer coils when, at the scales much exceeding the atomic ones, only the hydrodynamic forces determine the polymer behavior. Even this seemingly simple situation is not fully described in the literature. For example, the theory of the screening of HI due to the presence of other coils in the solution should be developed. The work [10] suggests that this screening is not only concentration dependent but is a time-dependent process. The aim of the present work was to give a description of this phenomenon and to find how it reveals in the observable quantities. As in the traditional theories, we started with the equations of motion for the test polymer, which should be solved together with the hydrodynamic equations for the solvent. The presented theory has the following limitations. The considered times are $t \gg R^2\rho/\eta$, where $R$ is the hydrodynamic radius of the coil. This means that the hydrodynamic memory effects [11, 20] (so far not observed in the polymer dynamics) are neglected. We are also restricted to $\theta$ solvents [2]; other cases require knowledge of the equilibrium distribution of the beads with the excluded volume interactions taken into account. Since only solutions of nonentangled polymers are considered, the studied concentrations of the chains are $c < 1/[\eta]$ [9]. Our approach differs from the previous bead-spring theories in several points. First, we do not *a priori* assume the validity of a concrete, Rouse or Zimm, model. Only the strength of the HI determines which type of the polymer behavior is dominant. Secondly, as distinct from the usual approximation leading to simple "universal" equations such as the $t^{1/2}$ (Rouse) or $t^{2/3}$ (Zimm) laws for the MSD of the polymer segments [2], the distribution of the internal modes of the polymer is not continuous. Within the usual approach it is not possible to describe the transitions between the Rouse and Zimm behavior of the polymer. However, this transition always takes place since at short times the HIs do not affect the polymer motion. The polymer moves according to the Rouse theory and at longer times, when the HIs develop, the regime of its dynamics changes to the Zimm one. The concept of the joint Rouse-Zimm model is essential also for the description of the polymer dynamics affected by other coils in the solution. Building the hydrodynamics of the solution of nonentangled polymers, we have shown that with the increase of the polymer concentration the Zimm contribution to the observable quantities (such as the coil diffusion coefficient or the viscosity of the solution) decreases and the polymer tends to behave (as distinct from the previous theories [2]) *exactly* in correspondence with the Rouse model. The same tendency is seen with the increase of time. Thus, the theory is able to explain the dynamic nature of the screening of HI. To take into account the presence of other coils in the solution, we have used the Brinkman's hydrodynamics for porous media, adopting it for the solvent flow in the solution where the obstacles to the flow are the polymer coils themselves. The main results of the presented approach consist in new equations for the position vector of the polymer beads and for a number of characteristics describing the behavior of flexible polymers in dilute solutions. These quantities could be verified in standard experiments such as the viscosimetry or light scattering, and in computer simulation studies similar to those in [10].

## Acknowledgements

We dedicate this paper to our untimely deceased colleague, Professor Alexandr Vsevolodovich Zatovsky.

This work was supported by the grants 1/3033/06 and 1/4021/07 from VEGA, Slovak Republic.

## Authors' information

[1]Department of Physics, Technical University of Kosice, Park Komenskeho 2, 042 00 Kosice, Slovakia
[2, 3]Institute of Physics, Faculty of Science, P. J. Safarik University, Jesenna 5, 041 54 Kosice, Slovakia